\documentclass[aps,twocolumn]{revtex4}
\usepackage{graphicx,amssymb}
\newcommand{\ket}[1]{\vert #1 \rangle}
\newcommand{\ketbra}[2]{\vert #1 \rangle \! \langle #2 \vert}
\newcommand{\Tr}[1]{\textrm{Tr}\,\! #1}
\begin{document}
\title{Photon statistics without counting photons}
\author{Andrea R. Rossi} \email{andrea.rossi@mi.infn.it}
\author{Stefano Olivares}\email{stefano.olivares@mi.infn.it}
\author{Matteo G. A.Paris} \email{matteo.paris@fisica.unimi.it}
\affiliation{Dipartimento di Fisica dell'Universit\`a degli Studi
di Milano, Italia}
\begin{abstract}
We show how to obtain the photon distribution of a single-mode field
using only avalanche photodetectors. The method is based on measuring
the field at different quantum efficiencies and then inferring the
photon distribution by maximum-likelihood estimation. The convergence
of the method and its robustness against fluctuations are illustrated
by means of numerically simulated experiments.
\end{abstract}
\date{\today}
\maketitle
\section{Introduction}
Optical signals and nonclassical states of light have been the
subject of constant attention over the last three decades. The
photon distribution, besides fundamental interest, plays a major
role in quantum communication schemes based on light beams, and
this stimulated many experiments on the statistical properties
of radiation \cite{dav}. More recently, the rapid development of
quantum information processing once again posed the issue of effective
methods to investigate photon statistics \cite{leb, mand}. Photon
distribution is usually obtained by photon counting in time intervals
much shorter than the coherence time of the beam under investigation.
A long time stability is thus needed and this requirement
becomes more and more strict as far as the intensity of light becomes
lower. As a matter of fact, most photon counting experiments involved
laser beams. Effective photon counters have been also developed, though 
their  current operating conditions are still extreme \cite{xxx}.
\par 
The advent of quantum tomography provided a novel
method to measure photon distribution \cite{mun,revt}. However, the
tomography of a state, which have been applied to several quantum
states \cite{raymerLNP}, needs the implementation of homodyne
detection, which in turn requires the appropriate mode matching of
the signal with a suitable local oscillator at a beam splitter.
\par 
In this paper, we address a simple method to obtain the
photon distribution without directly counting photons. In this scheme,
repeated preparations of the signal are revealed through avalanche
photodetectors (APD) at different quantum efficiencies. The resulting on/off
statistics is then used to reconstruct the photon distribution through
maximum-likelihood estimation. Since the model is linear and the photon
distribution is a set of positive numbers, then the maximum of the likelihood
functional can be found iteratively by the expectation-maximization (EM)
algorithm \cite{EM:alg:1,EM:alg:2}. The method does not require long time 
stability and involves
only simple optical components. The number of experimental runs
depends on the signal under investigation, roughly increasing with
its nonclassicality. 
\par
The idea of inferring photon distribution through detection at different
efficiencies has been already analyzed theoretically \cite{mogy}, and 
implemented to realize a multichannel fiber loop detector \cite{olom}.
Here we describe other possible implementations, analyze the reconstruction 
when only a subset of values $0<\eta_{\rm min} < \eta < \eta_{\rm max}<1$ 
of the quantum efficiency is available, and discuss in details the statistical 
properties of the method: convergence and robustness against fluctuations.
\par 
The paper is structured as follows: In Section
\ref{s:sta} we introduce the problem, and show that simple
estimation by inversion cannot be used due to lack of precision.
Possible implementations of the measurement scheme are also described.
Then, in Section \ref{s:ml}, we illustrate the reconstruction of
photon distribution by iterative solution of maximum likelihood
estimation. In Section \ref{s:exp} the convergence properties
of the method, as well as its robustness to fluctuations, are
discussed on the basis of several Monte Carlo simulated
experiments, performed on different kind of signals. Section
\ref{s:out} closes the paper with some concluding remarks.
\section{Estimation by inversion}\label{s:sta}
\begin{figure}[h]
\begin{center}
\includegraphics[width=0.3\textwidth]{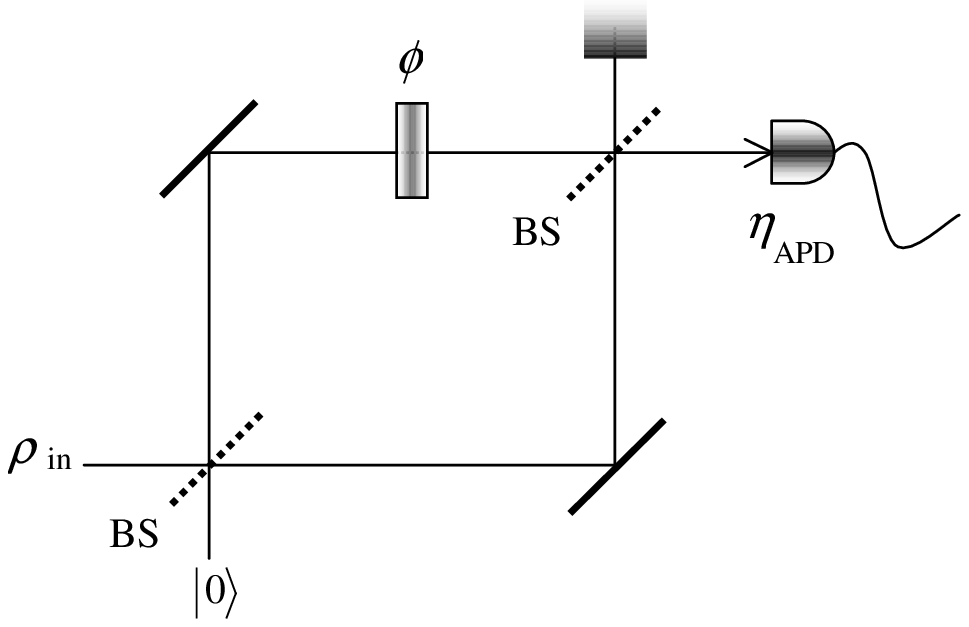}
\end{center}
\caption{A possible experimental setup for simulating different
quantum efficiencies. The signal passes through an
interferometer with internal phase-shift $\phi$ and is then revealed
by a high-efficiency APD. The transmissivity of the interferometer
is $\tau=\cos^2\phi$, and the overall efficiency of photodetection
is $\eta=\tau\eta_{\rm APD}$.} \label{f:fig0}
\end{figure}

Given a single-mode state $\varrho = \sum_{n,m} \varrho_{nm}
\ketbra{n}{m}$ we are interested in the photon distribution, {\em i.e}
in the set of positive numbers $\varrho_n \equiv \varrho_{nn} \geq 0$.
We assume to have at disposal APDs, which
can only discriminate the vacuum from the presence 
of radiation, with a certain quantum efficiency. This kind
of measurement, on/off photodetection, is described by the following
probability operator-valued measure (POVM) $\{\Pi_{\rm{off}},
\Pi_{\rm{on}}\}$
\begin{eqnarray} \label{povm}
\Pi_{\rm{off}}(\eta)&=&\sum_{n=0}^{\infty}(1-\eta)^n\ketbra{n}{n} \nonumber \\
\Pi_{\rm{on}}(\eta)&=&\mathbb{I}-\Pi_{\rm{off}}
\end{eqnarray}
$\eta$ being the detector's quantum efficiency, {\em i.e.} the
probability that an incoming photon lead to a {\em click}
of the detector. For any given state the detector does not click
with a probability $p_{\rm{off}}(\eta)=\Tr\{\rho \,
\Pi_{\rm{off}}(\eta)\}$, that reads as follows
\begin{eqnarray} \label{p:off}
p_{\rm{off}}(\eta)&=&\sum_{n=0}^{\infty}(1-\eta)^n\: \varrho_n
\:.
\end{eqnarray}
From now on we suppress the subscript ``off'' and
always mean $p_{\rm{off}}$ when we write $p$. The ``off''
probabilities for a set of $N$ detectors measuring the same
quantum state with different quantum efficiencies are then
\begin{eqnarray} \label{system}
\left\{ \begin{array}{lcr}
p_0(\eta_0)&=&\sum_n (1-\eta_0)^n \: \varrho_n  \\
p_1(\eta_1)&=&\sum_n (1-\eta_1)^n \: \varrho_n  \\
\phantom{i} & \vdots & \phantom{i} \\
p_{N}(\eta_{N})&=&\sum_n (1-\eta_{N})^n \: \varrho_n
\end{array}
\right.\label{sys}\:.
\end{eqnarray}
If we know all of the $\eta_\nu$'s values, equation (\ref{system})
is a linear system with unknowns $\left\{\varrho_n\right\}$.
In practice, it is not necessary to have at disposal many detectors
with different quantum efficiencies, since a suitable tuning of
$\eta$ can be obtained by optical filters or through an 
interferometric setup. In fact, besides the fiber loop scheme 
of \cite{olom}, different quantum efficiencies can be obtained 
inserting a set of optical filters before the detector, or by the scheme 
of Fig. \ref{f:fig0} where a single APD is needed, and 
lower efficiencies are obtained by varying the internal phase-shift 
$\phi$ of the interferometer. Since the overall transmissivity of 
the interferometer is $\tau=\cos^2\phi$, and a tuning of $\phi$ of 
the order of $\pi/500$ can be actually achieved, we may safely 
assume that about $100$ different values of $\eta=$ between 
$\eta_{\rm min}\simeq0$ and $\eta_{\rm max}=\eta_{\rm APD}$ can be 
obtained.
\par
Suppose now that the $\varrho_n$'s are negligible for
$n>\overline{n}$ and that we are able to measure the signal with
$N=\overline{n}$ different $\eta$'s. In this case equation
(\ref{system}) is a linear system of the form
\begin{eqnarray}
\mathbf{p}=\mathbb{V} \cdot \mbox{\boldmath$\varrho$}
\end{eqnarray}
where
\begin{eqnarray}
\mathbf{p}&=&\left\{p_0, p_1, \cdots, p_{\overline{n}-1} \right\}
^{T} \label{pn}\;
\\
\mbox{\boldmath$\varrho$}&=&\left\{\varrho_0, \varrho_1, \cdots,
\varrho_{\overline{n}-1} \right\} ^{T} \label{rhon}\;
\end{eqnarray}
and the coefficients matrix $\mathbb{V}$ (for $\eta_i \neq \eta_j$
$\forall \, i,j$) is a nonsingular Vandermonde matrix of order
$\overline{n}$. If we put $x_i=1-\eta_i$ the $\mathbb{V}$ matrix
reads
\begin{eqnarray} \label{vandermatrix}
\mathbb{V}=\left[ \begin{array}{ccccc} 1 & x_0 & x_0^2 & \cdots
&x_0^{\overline{n}-1} \\
1 & x_1 & x_1^2 & \cdots &
x_1^{\overline{n}-1} \\
\vdots & \vdots & \vdots & \ddots & \vdots \\
1 & x_{\overline{n}-1} & x_{\overline{n}-1}^2 & \cdots &
x_{\overline{n}-1}^{\overline{n}-1}
\end{array} \right]\:,
\end{eqnarray}
and the photon distribution can be obtained by matrix inversion
$\mathbf{\varrho}= \mathbb{V}^{-1}\: \mathbf{p}$.
The same approach can be used also when we are able to reveal the
state with $N>\overline{n}$ number of $\eta$'s. In this case the system
in Eq. (\ref{sys}) should be solved in the least square sense leading
to $\mathbf{\varrho}= \mathbb{W}\: \mathbf{p}$, where
$\mathbb{W}= (\mathbb{V}^T\mathbb{V})^{-1}\mathbb{V}^T$ is the
Moore-Penrose inverse of $\mathbb{V}$.\par
Unfortunately, the reconstruction of $\varrho_n$ by matrix
inversion cannot be used in practice since it would require
an unreasonable number of experimental runs.
In fact, most of the quantities $x_i$ entering the expression of
$\mathbb{V}$ are of order $10^{-1}$, and therefore the $\mathbb{V}^{-1}$'s
entries in the $j$-th line are of order $x^{-(j-1)}$.
This means that the reconstruction of $\varrho_j$ requires the
multiplication of the experimental frequencies $p_i$
by quantities of the order $x^{-(j-1)}$, which in turn implies
that a sound reconstruction of $\varrho_j$ needs that
$p_i$ must be precise at least to the $(j-1)$-th decimal digit, {\em
i.e.} a minimum of $10^{j-1}$ experimental runs.
In addition, for increasing $N$ and $\overline{n}$ the inversion of
(\ref{vandermatrix}) must be done numerically leading to errors
that quickly become unacceptably large.
\section{Maximum-likelihood estimation} \label{s:ml}
The problems illustrated in the previous Section can be circumvented
by considering equation (\ref{p:off}) as a statistical model for
the parameters ${\varrho_n}$ to be solved by maximum-likelihood
(ML) estimation. We assume $N>\overline{n}$ and, for sake of simplicity,
we define
\begin{eqnarray} \label{def}
p_{\nu}   &\equiv& p_{\nu}(\eta_\nu)\,, \nonumber\\
A_{\nu n} &\equiv& (1-\eta_{\nu})^n\,,
\end{eqnarray}
so that equations (\ref{system}) can be rewritten as
\begin{equation}
p_{\nu} = \sum_{n} A_{\nu n} \varrho_n\,.
\end{equation}
Since the model is linear and the parameters to be estimates are positive
(LINPOS problem), then the solution
can be obtained using the Expectation-Maximization algorithm (EM)
\cite{EM:alg:1, EM:alg:2}. By imposing the restriction
$\sum_n \varrho_n = 1$, we obtain the iterative solution
\begin{equation} \label{iterations}
\varrho_n^{(i+1)} = \varrho_n^{(i)}\sum_{\nu} \frac{A_{\nu
n}\,{\rm h}_{\nu}}{C_{\nu}\, p_{\nu}[\{\varrho_n^{(i)}\}]}
\end{equation}
where $\varrho_n^{(i)}$ is the value of $\varrho_n$ evaluated at
$i$-th iteration, $C_{\nu} = n_{\nu} \sum_m A_{\nu m}$, $n_{\nu}$
being the total number of experimental runs with $\eta =
\eta_{\nu}$, ${\rm h}_{\nu}$ is the number of {\em no-click}
events for $\eta = \eta_{\nu}$ and $p_{\nu}[\{\varrho_n^{(i)}\}]$
are the frequencies $p_{\nu}$ calculated using the reconstructed
distribution $\{\varrho_n^{(i)}\}$ at the $i$-th iteration. By introducing
the symbol $f_{\nu} = {\rm h}_{\nu} / n_{\nu}$ for the experimental
frequencies, the expression of $\varrho_n^{(i+1)}$ rewrites as
\begin{equation}
\varrho_n^{(i+1)} = \varrho_n^{(i)}\sum_{\nu} \frac{A_{\nu
n}}{\sum_m A_{\nu m}}
\frac{f_{\nu}}{p_{\nu}[\{\varrho_n^{(i)}\}]}\:.
\label{ems}\:.
\end{equation}
EM algorithm is known to converge unbiasedly to the ML
solution. Indeed, it has been already used to infer the photon
distribution from random phase homodyne data \cite{konrad}.
The confidence interval on the determination of the element
$\varrho_n$ can be given in terms of the variance $\sigma_n
=1/\sqrt{N\,F_{n}}$, $N$ being the number of 
measurements and $F_{n}$ the Fisher's
information \cite{cramer}
\begin{eqnarray} \label{fisher}
F_n=\sum_{\nu} \frac{1}{q_{\nu}} \left( \frac{\partial
q_{\nu}}{\partial \varrho_n} \right)^2\:,
\end{eqnarray}
where 
$$ 
q_\nu = \frac{p_\nu}{\sum_\nu p_\nu} = \frac{\sum_n A_{\nu n}\varrho_n}{
\sum_{\nu n} A_{\nu n} \varrho_n}\:,
$$
are the renormalized probabilities of no-click with quantum efficiency
$\eta_\nu$. $N_0=\sum_{\nu n} A_{\nu n} \varrho_n$ is the global fraction
of no-click events (irrespective of the quantum efficiency).
\par
Notice that Eq. (\ref{ems}) provides a solution once an initial 
distribution $\{ \varrho_n^{(0)}\}$ is chosen. 
In our simulated experiments we start from the uniform distribution
$\varrho_n^{(0)}=(1+\bar{n})^{-1}$  in $[0,\bar{n}]$. Other choices, 
the only constraint being $\varrho_n^{(0)}\neq0$, $\forall n$, do not 
dramatically influence the convergence properties of the algorithm.
\section{Monte Carlo simulated experiments and discussion}
\label{s:exp}
\begin{figure}[h]
\begin{center}
\includegraphics[width=0.4\textwidth]{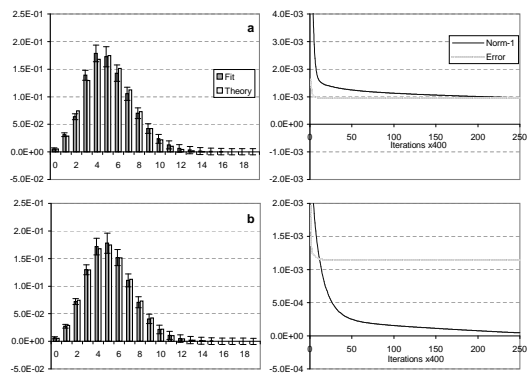} \label{coherent}
\end{center}
\caption{Reconstruction of the photon distribution of a coherent
state $|\alpha\rangle$ with average number of photons
$\langle a^{\dag}a \rangle\equiv |\alpha|^2= 5.20$. On the left: the
reconstructed photon distribution after $n_{\rm it}=10^4$ iterations.
On the right: the normalization factor $S^{(k)}$ and total error
$\varepsilon^{(k)}$ (\ref{error}) as a function of number of
iterations. The confidence interval has been evaluated as
$1/\sqrt{n_xF_n}$ where $F_n$ is the Fisher information.
The number of simulated data and the number of iterations 
are given by $n_x=n_{\rm{it}}=10^5$. In all the simulated 
experiments $N=50$ different quantum efficiencies have been used
with a minimum efficiency $\eta_{\rm{min}}=0.02$. The 
maximum efficiency is given by: (a) $\eta_{\rm{max}}=0.99$; 
(b) $\eta_{\rm{max}}=0.5$. The Hilbert space is truncated at 
$\overline{n}=20$.} \label{f:fig1}
\end{figure}

\begin{figure}[h]
\begin{center}
\includegraphics[width=0.4\textwidth]{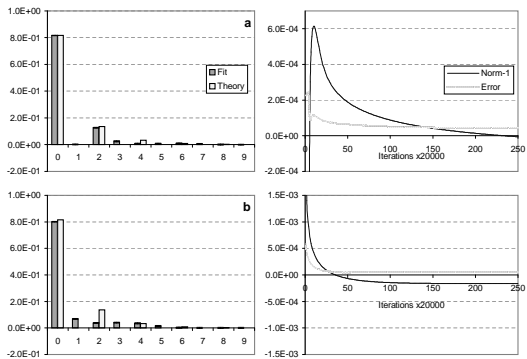}
\end{center}
\caption{Reconstruction of the photon distribution of a squeezed
state with squeezed fraction $\zeta=0.99$ and average photon
number $\langle a^{\dag}a \rangle=0.5$. Notice that a larger number of
iterations is needed in comparison with the coherent signals'
case. The number of simulated data and the number of iterations 
are given by $n_x=10^5$,$n_{\rm{it}}=5\times10^5$. The maximum efficiency 
is given by: (a) $\eta_{\rm{max}}=0.99$; (b) $\eta_{\rm{max}}=0.7$. 
The other parameters are the same as in Fig. \ref{f:fig1}.} \label{f:fig2}
\end{figure}

\begin{figure}[t]
\begin{center}
\includegraphics[width=0.4\textwidth]{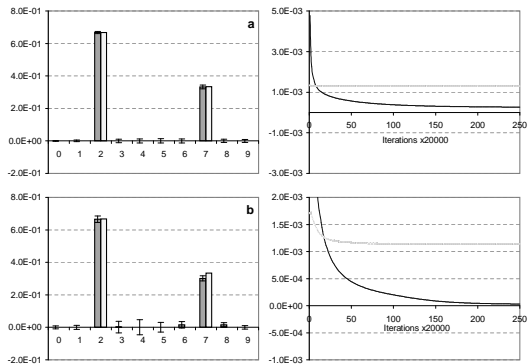}
\end{center}
\caption{Reconstruction of the photon distribution of
a unbalanced superpositions of number states (see text). 
The number of simulated data
and the number of iterations are given by:
(a) $n_x=10^4$,$n_{\rm{it}}=10^6$; The maximum efficiency 
is given by: (a) $\eta_{\rm{max}}=0.99$; (b) $\eta_{\rm{max}}=0.5$. 
The other experimental parameters are the same as in Fig. \ref{f:fig1}.}
\label{f:fig3}
\end{figure}

We have performed several numerical simulations in order to check
the accuracy and reliability of our method by varying the different
parameters. Since our solution of the ML estimation is obtained from
an iterative solution, the most important aspect to keep under control
is its convergence. As a measure of convergence we use the total
absolute error at the $k$-th iteration
\begin{eqnarray} \label{error}
\varepsilon^{(k)} =\sum_{\nu=0}^N |\epsilon_\nu^{(k)}|\:,
\end{eqnarray}
where
\begin{eqnarray}
\epsilon_\nu^{(k)}=
p_\nu- p_\nu [\{ \varrho_n^{(k)}\}]=
p_\nu- \sum_{n=0}^{\overline{n}-1}(1-\eta_\nu)^n \varrho_n^{(k)}\:.
\end{eqnarray}
The total error $\varepsilon^{(k)}$ measures the distance 
of the probabilities $p_\nu [\{ \varrho_n^{(k)}\}]$, as 
calculated at the $k$-th iteration, from the actual probabilities
as calculated from the theoretical photon distribution.
As a measure of accuracy we adopt the fidelity
\begin{eqnarray}
G^{(k)} = \sum_n \sqrt{\varrho_n \: \varrho_n^{(k)}}
\label{fid}\;
\end{eqnarray}
between the reconstructed distribution and the theoretical one. In
Figs. \ref{f:fig1}-\ref{f:fig3} (right) we report $\varepsilon^{(k)}$
versus the number of iterations for different signals.
As it is apparent from the plots, the total error is a good marker
for the convergence of the algorithm, while the normalization factor
$S^{(k)}=\sum_n \varrho_n^{(k)}-1$ (ideally zero at each step)
is not. Notice, however, that the minimum of the total error does
not always coincide with the maximum fidelity of reconstruction
(Figs. \ref{f:fig5} and \ref{f:fig6}), which means that our method is
slightly biased, especially for fast reconstruction, {\em i.e.}
when it converges quickly as it happens for coherent signals. We
have numerically observed that this problem can be circumvented by
using a number of iterations $n_{\rm it} \simeq n_x$ approximately
equal to the number of data. Currently we have no precise
explanation of this phenomenon, and provide it as a heuristic
prescription leading to best performances for a large class of
quantum signals.
\par
We have performed simulated experiments for coherent states
$|\alpha\rangle=D(\alpha)|0\rangle$, squeezed states $|\alpha,
\xi\rangle=D(\alpha)S(\xi)|0\rangle$ and superposition of Fock
states $|\psi\rangle=2^{-1/2} (|n_1\rangle+|n_2\rangle)$, where
$D(\alpha)=\exp(\alpha a^\dag - \bar\alpha a)$ is the displacement
operator and $S(\xi)=\exp(\frac12 \xi a^{\dag 2} - \frac12 \bar \xi a^2)$
is the squeezing operator. Squeezed states have been parametrized
through the total average photon number and the squeezing fraction
\begin{eqnarray}
\langle a^\dag a\rangle &=& |\alpha|^2 + |\xi|^2/(1-|\xi|^2) \nonumber \\
\zeta &=& 1-|\alpha|^2/\langle a^\dag a\rangle\:.
\label{parsq}\;
\end{eqnarray}
The value $\zeta=1$ corresponds to a squeezed vacuum and $\zeta=0$
to a coherent state. As shown in Fig. \ref{f:fig1}, the algorithm
converges quite fast for coherent states, while for nonclassical
states such as squeezed states (Fig. \ref{f:fig2}) the number of
needed iterations is larger. The right plots in Fig. \ref{f:fig2}
also indicate that increasing $N$ does not always improve
accuracy. In Fig. \ref{f:fig3} we show reconstruction for
the unbalanced superpositions of Fock states 
$|\psi_2\rangle=(2/3)^{1/2} |2\rangle+(1/3)^{1/2}|7\rangle$.
\par
Concerning the values of the quantum efficiency, we used 
$N$ values of $\eta$ uniformly distributed in $[\eta_{\rm min},
\eta_{\rm max}]$ with $\eta_{\rm min}\simeq 0$ and $\eta_{\rm max} < 1$.
In principle, a different distribution (not uniform) may influence 
the performances of the
algorithm. We found, however, that both  convergence and accuracy
are not much affected by a different choice, which may become
relevant only if the spacing between the efficiency values becomes
smaller. It should be noticed that the algorithm works well also 
when $\eta_{\rm max}$ is considerably smaller than unit.
This is a relevant feature of the method in view of its 
experimental implementations in different working regimes.
In Figs. \ref{f:fig1}-\ref{f:fig3} (bottom left) we report 
the reconstructions obtained assuming $\eta_{\rm max}=0.5$ 
(coherent states and superpositions) and $\eta_{\rm max}=0.7$ 
(squeezed states).
\par
In experiments where we have no {\em a priori} information on the
state under investigation it could happen that part, or even most,
of the number distribution $\varrho_n$ lies outside the
reconstruction region (from $0$ to $\bar n$). In this case we have
checked that the algorithm is able to reconstruct accurately the
norm of the included part, such that a simple check of the
distribution norm allows to optimize $\bar n$ (and in turn $N$) in
few steps. This is a remarkable feature of the algorithm, since in
general a large $\bar n$ improves convergence but doesn't
guarantee better accuracy.
\par
\begin{figure}[h!]
\begin{center}
\includegraphics[width=0.4\textwidth]{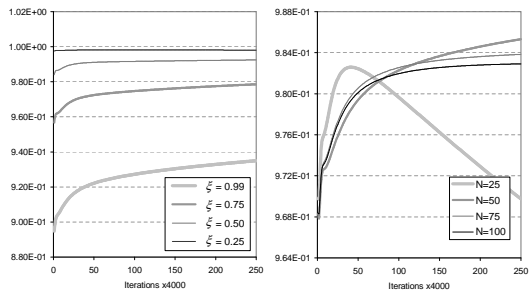}
\end{center}
\caption{Fidelity $G^{(k)}$ versus the number of iterations.
Left: for a squeezed state with $\langle a^{\dag}a \rangle=1.0$ and different
squeezing fractions $\zeta$ (left). Right: for a squeezed state
with $\langle a^{\dag}a \rangle=1.0$, $\zeta=0.75$ and different numbers
$N$ of $\eta$'s values. In both cases the maximum number of iterations
is $n_{\rm{it}}=10^6$.
} \label{f:fig4}
\end{figure}
The error bars in the plots have been calculated using the Fisher
information (\ref{fisher}), that explicitly reads:
\begin{eqnarray} \label{fisherexp}
F_n&=&\frac{1}{{N_0}^3}\sum_{\nu}\frac{1}{p_\nu} 
\left( A_{\nu n}{N_0}- p_\nu \sum_k A_{k n}  \right)^2,
\end{eqnarray}
where ${N_0}$ is the total number of no click events.
\begin{figure}[h!]
\begin{center}
\includegraphics[width=0.35\textwidth]{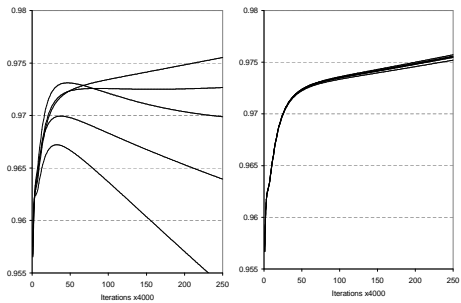}
\end{center}
\caption{Fidelity $G^{(k)}$ versus the number of iterations for a
squeezed state with $\langle a^{\dag}a \rangle=1.5$ and $\zeta=0.75$. Each line
represent a different simulated run. Left: $n_x=10^5$, right: $n_x=10^6$.
In both the plot the maximum iterations number is $n_{\rm{it}}=10^6$,
while the other parameters are the same as in Fig. \ref{f:fig1}.}
\label{f:fig5}
\end{figure}
\par
A question may arise about the robustness of the method against
fluctuations in the value of the $\eta_\nu$ (which, in the case of
the interferometric implementation of Fig. \ref{f:fig0}, may occur
as a consequence of {\em phase} fluctuations), {\em i.e.} whether or
not their precise knowledge is needed. In order to check robustness we 
have performed simulated experiments where, during the run, 
the quantum efficiency may fluctuate. In particular, we assumed each 
$\eta_\nu$ uniformly distributed in the range 
$(-\sigma+\eta_\nu,\sigma+\eta_\nu)$, 
where
\begin{eqnarray} \label{sigma}
\sigma=\frac{\eta_{\rm{max}}-\eta_{\rm{min}}}{a\,N}\:, 
\end{eqnarray}
and $a$ is a positive number. The value $a=2$ corresponds to 
each $\eta_\nu$ fluctuating in an interval as large as the spacing 
$(\eta_{\nu+1}-\eta_\nu)$ around its expected value. The values of $p_{\nu}$ change
accordingly during the run. Our results are summarized in
Fig. \ref{f:fig6}. The reconstruction is not dramatically 
affected by fluctuations, though errors bars are slightly larger. 
We conclude that the method is robust against fluctuations.
\par
\section{Conclusions}
\label{s:out}
We analyzed in details an iterative algorithm to infer the photon
distribution of a single-mode radiation field using only avalanche
photodetectors. The method is accurate and statistically reliable
for a large class of Gaussian (coherent and squeezed) and
non Gaussian states (superpositions and mixtures of $\ket{n}$ states),
provided that on/off photodetection may be performed at different
quantum efficiencies. The scheme involves only simple optical 
components, and allows reconstruction with APD quantum efficiency 
considerably smaller than unit. The convergence of the 
method, and its robustness against fluctuations of quantum efficiency 
have been demonstrated numerically, by means of Monte Carlo simulated 
experiments.
\begin{figure}[h]
\begin{center}
\includegraphics[width=0.35\textwidth]{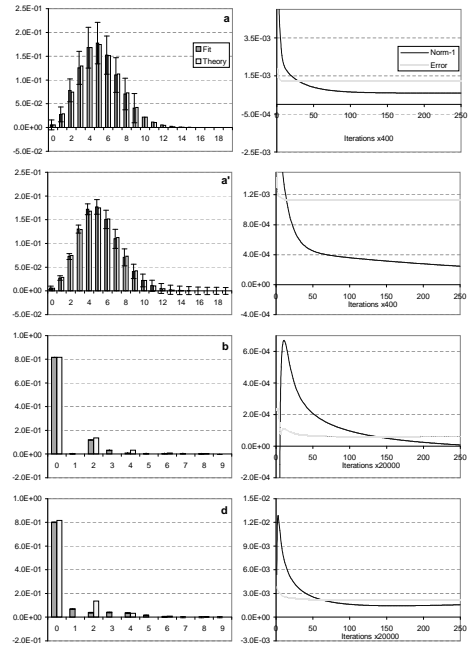}
\end{center}
\caption{Reconstruction of the photon distribution for
different signals for fluctuating $\eta_\nu$. (a,a'): coherent state
with $\langle a^{\dag}a \rangle=5.20$; (b,b'): squeezed state with 
$\langle a^{\dag}a \rangle=0.50$
and $\zeta=0.99$. The number of simulated data and the number of iterations
are given by:
(a,a')$n_x=10^5$,$n_{\rm{it}}=10^5$; (b,b') $n_x=10^6$,$n_{\rm{it}}=
5\times10^6$,$\langle a^{\dag}a \rangle_{\rm{fit}}=5.01$. In all the simulated 
experiments we set $a=2$ in Eq. (\ref{sigma}).
The maximum efficiency 
is given by: (a,b) $\eta_{\rm{max}}=0.99$; (a') $\eta_{\rm{max}}=0.5$; 
(b') $\eta_{\rm{max}}=0.7$. 
The other experimental parameters are the same as in Fig. \ref{f:fig1}.}
\label{f:fig6}
\end{figure}
\section*{Acknowledgments}
We thank M. Bondani and A. Ferraro for reading of the manuscript, and 
Z. Hradil for pointing out relevant references, and for his friendly 
suggestions. MGAP thanks Marco Genovese for a fruitful discussion. 

\end{document}